# PHOTOPOLARIMETRICAL STUDY OF BLAZAR-TYPE AGN OJ287 IN 2012-2015 WITH THE 2M RCC TELESCOPE AT NAO ROZHEN

VLADIMIR BOZHILOV[1], EVGENI OVCHAROV[1], MILEN MINEV[1,2,3], YORDAN DARAKCHIEV[1], ANGEL DIMITROV[1], STEFAN GEORGIEV[1,2], MANOL GERUSHIN[1], BORISLAV SPASSOV[1,2], KALINA STOIMENOVA[1]

[1] *Department of Astronomy, Faculty of Physics, Sofia University "St. Kliment Ohridski", 5 "James Bourchier" Blvd., 1164 Sofia, Bulgaria*
[2] *Institute of Astronomy and National Astronomical Observatory, Bulgarian Academy of Sciences, 72, Tsarigradsko Chaussee Blvd. 1784, Sofia Bulgaria*
[3] *Institute for Nuclear Researches and Nuclear Energy, 72, Tsarigradsko Chaussee Blvd. 1784, Sofia, Bulgaria*

*ВЛАДИМИР БОЖИЛОВ, ЕВГЕНИ ОВЧАРОВ, МИЛЕН МИНЕВ, ЙОРДАН ДАРАКЧИЕВ, АНГЕЛ ДИМИТРОВ, СТЕФАН ГЕОРГИЕВ, МАНОЛ ГЕРУШИН, БОРИСЛАВ СПАСОВ, КАЛИНА СТОИМЕНОВА, ФОТОПОЛЯРИМЕТРИЧНО ИЗСЛЕДВАНЕ НА БЛАЗАРА OJ287 В ПЕРИОДА 2012-2015 Г. С 2M RCC ТЕЛЕСКОП НА НАО-РОЖЕН*

Представяме резултатите от фотополяриметрично изследване на блазара OJ287 в периода ноември 2012 — януари 2015 г. Налюденията са извършени с помощта на фокалния редуктор FoReRo-2 на 2-метровия RCC телескоп на Националната астрономична обсерватория (НАО) Рожен. Наблюдаваната промяна в позиционния ъгъл (P.A.) отговаря на въртене на равнината на поляризация с 6,23 ±0,05 градуса/ден. Открита е индикация за корелация между кривата на блясъка на обекта във филтър R и промяната в степента на поляризация.

*VLADIMIR BOZHILOV, EVGENI OVCHAROV, MILEN MINEV, YORDAN DARAKCHIEV, ANGEL DIMITROV, STEFAN GEORGIEV, MANOL GERUSHIN, BORISLAV SPASSOV, KALINA STOIMENOVA,* PHOTOPOLARIMETRICAL STUDY OF BLAZAR-TYPE AGN OJ287 IN 2012-2015 WITH THE 2M RCC TELESCOPE AT NAO ROZHEN

We present the results of a photopolarimetric study of the blazar OJ287 in the period November 2012 – January 2015. Observations were conducted using the Focal Reductor FoReRo-2 of the 2-meter RCC telescope of the National Astronomical Observatory (NAO) – Rozhen. The observed change of the position angle (P.A.) corresponds to mean rotation of the plane of polarization of 6.23 ±0.05 deg/day, in good agreement with previous measurements. An indication of a correlation between the change of brightness in R-band and the change in the degree of polarization is also observed.

*For contact*: Dr. Vladimir Bozhilov, Department of Astronomy, Faculty of Physics, Sofia University "St. Kliment Ohridski", 5 "James Bourchier" Blvd., 1164 Sofia, Bulgaria, phone: +35928161413, email: *vbozhilov@phys.uni-sofia.bg*

## 1. INTRODUCTION

Blazars are amongst the most variable and energetic objects in the universe. They are a type of active galactic nuclei (AGN), powered by matter, falling on a supermassive black hole in the center of a large galaxy. OJ287 is a BL Lac type AGN at z=0. 306. It is one of the most widely observed and well-studied objects of this type. Observational data dates back to 1890 and the analysis gives a confirmed period of outbursts at intervals of roughly every 11 years. [1]. The most prominent theoretical model for OJ287 is the one of a binary black hole (BBH) system in which the smaller black hole crosses the accretion disk of the larger one [2]. OJ287 exhibits some characteristics, which are common for all blazars of such type. However, the periodicity of its outbursts is an unusual behavior [3]. Even the best current models fails to predict some of the outbursts that are observed and polarization behavior might be the key to determine which models work better. The observations of OJ287 show linear polarization at multiple wavelengths [4]. According to polarimetrical studies, a significant wavelength dependence in the degree of polarization is also observed [5]. Observations in the optical and radio- wavelengths indicate strong variability in the degree of polarization (P) and position angle (P.A.) [6].

## 2. OBSERVATIONS

This study presents our original polarimetrical and photometrical observations of OJ287 blazar-type AGN. We gathered and analyzed data from the nights of 17-18 Nov 2012, 13 Jan 2013, 04-05 Apr 2013, 30-31 Mar 2013, 19-21 Oct 2014, 17 Nov 2014, 13-14 Jan 2015. The observations were conducted with the 2-m RCC telescope at the National Astronomical Observatory Rozhen, Bulgaria, equipped with the focal reducer FoReRo2 [7]. The observation data for November 2012 through January 2015 is given in **Table 1**. Please note, that part of the data in the period November 2012 – April 2013 was published in [13] and [14], but the data from 2014-2015 is published here for the first time.

For the observations we use a colour splitter that transmits redder than 5800 Å light into the red channel and reflects bluer than 5100 Å into the blue channel of the reducer. We used an optical element with two combined Wollaston prisms that form a single polarizer. The P.A. of the prism differ by 45 deg, so we get four polarized beams with orientations at 0, 45, 90 and 135 deg each [13].

With this setup and by using the Stokes equations ([8]), we perform the polarization determinations. Description of the method of measurements and the calculations can be found in [9] and [13].

**Table 1.** OJ287 observational data for 2012 November - 2015 January.

| Object | JD at midnight – 2456700 [d] | Images | Exposure [s] | Total integration time [s] |
|---|---|---|---|---|
| HD 10476 | 49.4 | 30 | 0.2 | 6 |
| HD 14433 | 49.4 | 30 | 0.5 | 15 |
| OJ287 | 49.5 | 30 | 60 | 1800 |
| OJ287 | 49.5 | 30 | 60 | 1800 |
| OJ287 | 49.6 | 30 | 60 | 1800 |
| OJ287 | 49.6 | 30 | 60 | 1800 |
| OJ287 | 49.6 | 30 | 60 | 1800 |
| OJ287 | 49.6 | 20 | 60 | 1200 |
| OJ287 | 50.5 | 10 | 60 | 600 |
| OJ287 | 50.5 | 10 | 60 | 600 |
| OJ287 | 50.5 | 10 | 60 | 600 |
| HD 90508 | 106.4 | 10 | 3 | 30 |
| HD 90508 | 106.5 | 10 | 1 | 10 |
| HD 43384 | 106.5 | 10 | 1 | 10 |
| OJ287 | 106.4 | 10 | 300 | 3000 |
| OJ287 | 106.5 | 10 | 300 | 3000 |
| HD 144287 | 187.5 | 5 | 0.5 | 2.5 |
| HD 144287 | 188.4 | 5 | 0.5 | 2.5 |
| HD 154445 | 187.5 | 5 | 0.1 | 0.5 |
| HD 154445 | 188.5 | 5 | 0.2 | 1 |
| OJ287 | 187.5 | 3 | 300 | 900 |
| OJ287 | 188.4 | 15 | 300 | 4500 |
| HD 43384 | 546.5 | 5 | 0.1 | 0.5 |
| HD 144287 | 546.5 | 5 | 0.2 | 1.0 |
| OJ287 | 546.5 | 5 | 120.0 | 600.0 |
| OJ287 | 547.5 | 25 | 150.0 | 3750.0 |
| HD 23512 | 749.5 | 10 | 0.5 | 5.0 |
| HD 65583 | 749.5 | 10 | 0.2 | 2.0 |
| OJ287 | 749.5 | 10 | 200.0 | 2000.0 |
| OJ287 | 750.5 | 10 | 200.0 | 2000.0 |
| HD 23512 | 778.5 | 10 | 0.2 | 2.0 |
| HD 65583 | 778.5 | 10 | 0.2 | 2.0 |
| OJ287 | 778.5 | 5 | 300.0 | 1500.0 |
| HD 23512 | 835.5 | 20 | 0.5 | 10.0 |
| HD 65583 | 835.5 | 20 | 0.1 | 2.0 |
| OJ287 | 835.5 | 30 | 100.0 | 3000.0 |
| OJ287 | 836.5 | 20 | 80.0 | 1600.0 |
| OJ287 | 836.5 | 10 | 100.0 | 1000.0 |

## 3. DATA REDUCTION

Polarization can be determined using the Stokes parameters I, Q, U and V. They fully describe the state of polarization of an arbitrary electromagnetic wave [10]. A complete derivation of these parameters, as well as guides and methods for measuring them, are given in [8], chapters 1.6 and 1.7.

The method for measurement of the Stokes parameters is given by [9]. This method requires the usage of standard stars whose P.A. and polarization, are well known. Furthermore, it requires that some stars are with low P, and some – with high P. In this article, we have used the low-polarization standard stars HD 144287 and HD 65583, and the high-polarization standards HD 43384 and HD 23512.

All images have been reduced using the software package IRAF (Image Reduction and Analysis Facility, [11]). The photometric measurements have been reduced using the widely-known and accepted procedure: correction for noise (thermal and other), "flat field" correction, and median combination of the resulting images. The method, its purpose, application, advantages and drawbacks, are described broadly in many articles, e.g. [12].

Each of the combined images is then subjected to aperture photometry with aperture radius approximately equal to the full width at half maximum (FHWM). The process for finding P and P.A. is described in [9] and [13].

## 4. RESULTS AND DISCUSSION

Measurements of the polarization and the P.A., as well as the number of complete rotations of the plane of polarization $k$ (see [13] for elaborate details) are shown in **Table 2**. Magnitude in R-band and polarization curves are shown simultaneously in **Figure 1.** Measurements of the P.A. with the applied fitting are shown in **Figure 2.**

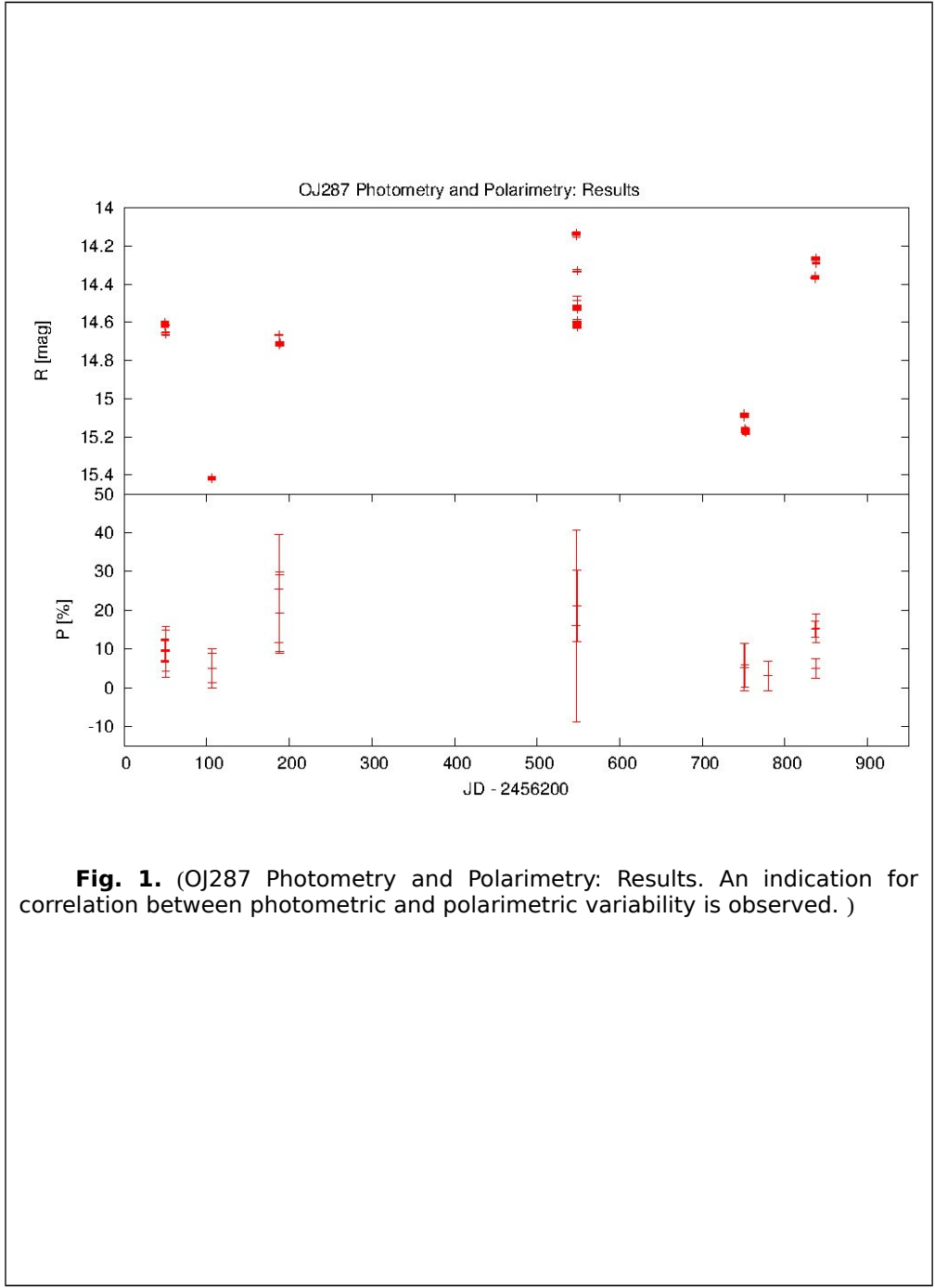

**Fig. 1.** (OJ287 Photometry and Polarimetry: Results. An indication for correlation between photometric and polarimetric variability is observed. )

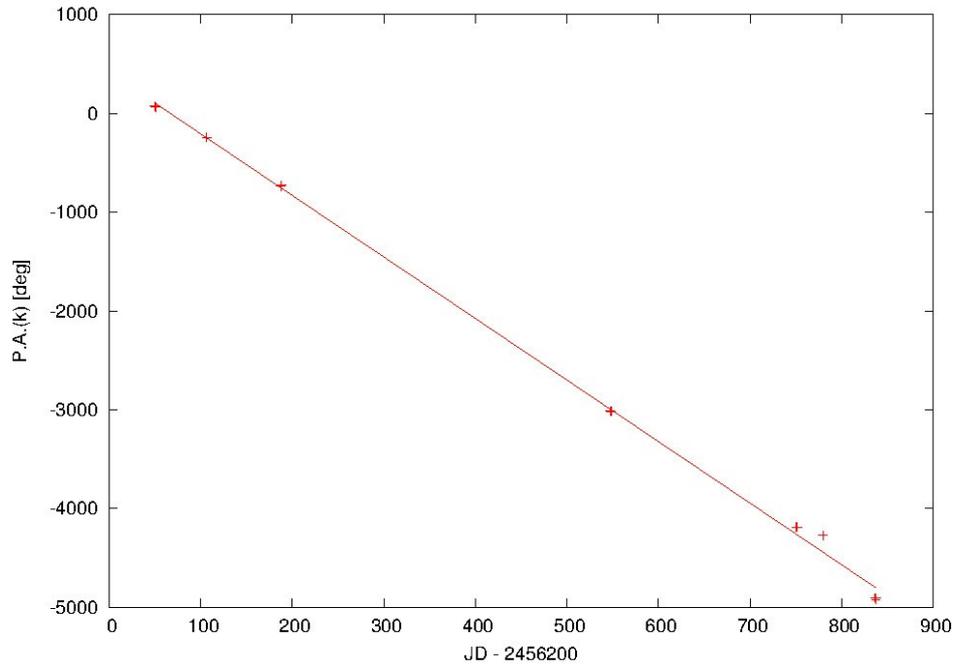

**Fig. 2.** (P.A. Measurements for OJ 287. Observed P.A. change corresponds to rotation of 6.23 (± 0.05) deg/day.)

Our results show mean rotation of the plane of polarization of 6.23 ±0.05 deg/day, which is in good agreement with previously measured values [13]. However, further studies are necessary to understand better the physical properties of OJ 287, especially in the periods following the flares of this blazar [15].

| DATE | JD - 2456200 | P [%] | Error [%] | P.A. [deg] | P.A.(k) [deg] | Error [deg] | k |
|---|---|---|---|---|---|---|---|
| 17 NOV 2012 | 49.495 | 9.80 | 2.72 | 73.29 | 73.29 | 3.15 | 0 |
| | 49.523 | 9.84 | 2.69 | 74.52 | 74.52 | 3.15 | 0 |
| | 49.552 | 9.71 | 2.67 | 74.77 | 74.77 | 3.15 | 0 |
| | 49.588 | 9.89 | 2.76 | 74.62 | 74.62 | 3.15 | 0 |
| | 49.618 | 9.45 | 2.78 | 73.53 | 73.53 | 3.16 | 0 |
| | 49.646 | 9.48 | 2.76 | 72.53 | 72.53 | 3.16 | 0 |
| 18 NOV 2012 | 50.491 | 9.27 | 6.56 | 62.60 | 62.60 | 3.37 | 0 |
| | 50.500 | 9.54 | 5.31 | 64.45 | 64.45 | 3.29 | 0 |
| | 50.510 | 9.53 | 5.34 | 64.17 | 64.17 | 3.29 | 0 |
| 13 JAN 2013 | 106.364 | 4.93 | 5.10 | 117.33 | -242.67 | 3.20 | 1 |
| | 106.496 | 5.07 | 3.91 | 117.91 | -242.09 | 3.07 | 1 |
| 04 APR 2013 | 187.448 | 25.59 | 13.90 | 163.65 | -736.35 | 1.53 | 2.5 |
| 05 APR 2013 | 188.348 | 19.37 | 10.52 | 172.34 | -727.66 | 1.53 | 2.5 |

Table 2. Photopolarimetrical results for Blazar-type AGN OJ287 in the period 2012-2015

|  | 188.385 | 19.35 | 9.91 | 172.37 | -727.63 | 1.52 | 2.5 |
| --- | --- | --- | --- | --- | --- | --- | --- |
| 30 MAR 2014 | 547.390 | 15.95 | 24.74 | 46.72 | -3013.28 | 6.52 | 8.5 |
| 31 MAR 2014 | 548.337 | 21.1 | 9.15 | 42.04 | -3017.96 | 5.94 | 8.5 |
| 19 OCT 2014 | 750.576 | 5.28 | 6.15 | 132.74 | -4187.26 | 3.58 | 12 |
| 20 OCT 2014 | 751.590 | 5.81 | 5.57 | 130.1 | -4189.90 | 3.47 | 12 |
| 17 NOV 2014 | 779.524 | 3.09 | 3.77 | 48.74 | -4271.26 | 2.04 | 12 |
| 13 JAN 2015 | 836.558 | 15.18 | 2.1 | 135 | -4905.00 | 2.41 | 14 |
| 14 JAN 2015 | 837.359 | 4.99 | 2.56 | 116.84 | -4923.16 | 2.59 | 14 |
| 15 JAN 2015 | 837.673 | 15.34 | 3.72 | 132.82 | -4907.18 | 2.46 | 14 |

## 5. CONCLUSIONS

Photopolarimetrical studies of the blazar OJ287 are a powerful tool to determine between different plausible theoretical models. It is particularly important to perform such measurements during outbursts when the characteristic double-peaked flares could be separated ([15]). The first peak is known to be of thermal origin while the second one could be due to either polarized synchrotron radiation or to unpolarized / low-polarized bremsstrahlung (braking radiation).

In this article we extend the polarimetrical and photometrical observations we began in the period of November 2012 through April 2013, described in [13] and [14]. We add more recent data obtained in the period from March 2014 through January 2015 in an effort to refine our results for the measurement of the rotation of the plane of polarization of OJ287. From the variation of the P.A. we were able to derive mean rotation of the plane of polarization of 6.23($\pm$ 0.05) deg/day, compared to the result from the previous data set of 5.80 ($\pm$ 0.03) deg/day [13]. The combined photopolarimetrical observations from the whole period suggest significant correlation between polarization and brightness in R-band which is an expected result in some of the theoretical models [13].

Further polarimetrical observations of this object are strongly encouraged, since such results are scarce, but can serve as an important constrain to the theoretical models.

**Acknowledgements.** The analysis of the gathered data and the preparation of this publication was partially supported by project „Time domain astrophysics of selected active galactic nuclei and Local group stars", funded by Contract DN18/10 with the National Science Fund of Bulgaria. Part of the work was done by students, participating in the lecture course „Preparation of Scientific Articles in Astronomy", read in the Dept. of Astronomy, Faculty of Physics, University of Sofia.

REFERENCES

[1] Sillanpaa, A., Haarala, S., Valtonen, M. J., Sundelius, B., Byrd, G. G.,Astrophysical Journal, Part 1 (ISSN 0004-637X), vol. 325, Feb. 15, 1988, p. 628-634
[2] Lehto, Harry J., Valtonen, Mauri J., Astrophysical Journal, 1996, v.460, p.207
[3] Takalo, L. O., Vistas in Astronomy, 1994, vol. 38, Issue 1, pp.77-109
[4] Kinman, J.D. et al., Astronomical Journal, 1974, Vol. 79, p. 349-357
[5] Sitko, M. L., Schmidt, G. D., & Stein, W. A., Astrophysical Journal Supplement Series (ISSN 0067-0049), vol. 59, Nov. 1985, p. 323-342.
[6] Pursimo, T. et al., Astronomy and Astrophysics Supplement, 2000, v.146, p.141-155
[7] Jockers K. et al., Kinematika Fiz. Nebesnykh Tel Suppl., 2000, 3, 13
[8] Landi Degl'Innocenti E., Landolfi M.,, Astrophysics and Space Science Library, 2004 , Vol. 307, Polarization in Spectral Lines. Kluwer, Dordrecht, p. 15
[9] Geyer E., Jockers K., Kiselev N. N., Chernova G. P., Ap&SS, 1996, 239, 269
[10] Walker, M.J.,, American J. Phys. 1954, 22, 170.
[11] Tody, D. 1986, The IRAF Data Reduction and Analysis System, Proc. SPIE Instrumentation in Astronomy VI, 627, 733.
[12] Massey, P., A User's Guide to CCD reductions with IRAF, Tucson, 1997.
[13] Bozhilov, V., Ovcharov, E., Nikolov, G., MNRAS, 2014, Vol. 439, Issue 1, p.639-643
[14] Bozhilov, V.; Borisov, G.; Ovcharov, E.; BgAJ, 2013, Vol. 19, p. 29
[15] Gupta, A. C. et al; MNRAS, 2017, Vol. 465, Issue 4, p.4423-4433.